# Testability of Reversible Iterative Logic Arrays

Avik Chakraborty

***Abstract*— Iterative Logic Arrays (ILAs) are ideal as VLSI sub-systems because of their regular structure and it's close resemblance with FPGAs (Field Programmable Gate Arrays). Reversible circuits are of interest in the design of very low power circuits where energy loss implied by high frequency switching is not of much consideration. Reversibility is essential for Quantum Computing. This paper examines the testability of Reversible Iterative Logic Arrays (ILAs) composed of reversible k-CNOT gates. For certain ILAs it is possible to find a test set whose size remains constant irrespective of the size of the ILA, while for others it varies with array size. Former type of ILAs is known as Constant-Testable, i.e. C-Testable. It has been shown that Reversible Logic Arrays are C-Testable and size of test set is equal to number of entries in cell's truth table implying that the reversible ILAs are also Optimal-Testable, i.e. O-Testable. Uniform-Testability, i.e. U-Testability has been defined and Reversible Heterogeneous ILAs have been characterized as U-Testable. The test generation problem has been shown to be related to certain properties of cycles in a set of graphs derived from cell truth table. By careful analysis of these cycles an efficient test generation technique that can be easily converted to an ATPG program has been presented for both 1-D and 2D ILAs. The same algorithms can be easily extended for n-Dimensional Reversible ILAs.**

## I. Introduction

Reversible circuits are classical counterparts of Quantum Circuits which are inherently reversible. It has manifold applications in low power CMOS quantum computing, nanotechnology, optical computing, computer graphics, DNA technology and cryptography. In quantum circuits, the unit of quantum information is a qubit, which can be either in a zero state ($|0>$) or a one state ($|1>$). It can also be in a state which is a superposition of these states, i.e. $\alpha|0> + \beta|1>$ where $\alpha$ and $\beta$ are complex numbers called amplitudes so that $|\alpha|^2 + |\beta|^2 = 1$. Quantum computation can solve exponentially hard problems (ex. Prime Factorization) in polynomial time by exploiting the superposition.

Reversible circuits have the property that every distinct input pattern yields a distinct output pattern, and the input and output wires are equal in number. Such circuits are of interest for several reasons. Reversible circuits are information-loss less and hence tend to dissipate relatively little energy. According to Landauer's principle [1, 2], it is possible to construct a computer using reversible circuits that can compute with arbitrarily small amounts of energy. Erasure of 1 bit information gives rise to entropy by $klog_e2$ and heat dissipation by $kT\ log_e2$ where $T$ is the temperature of the environment. Here $k$ is boltzmann's constant. Reversibility is an essential property of the circuits needed for quantum computation [3], which is a major motivation behind this study.

The testing of ILAs has been widely studied in the past and even more so in recent times due to advances in VLSI which have made these structures attractive to the circuit designer. Most test generation techniques use the regular cell interconnection structure of the ILAs in one way or the other [4, 5, 6, 7, 8]. Stroud etal [9] have used concept of ILA testing for a BIST approach to FPGA testing.

In this paper we examine testability of Iterative Logic Arrays (ILA's) constructed from a library of reversible gates called *k*-controlled-NOT (*k*-CNOT) gates. A *k-CNOT gate* has $k + 1$ input wires and $k + 1$ output wires. It transmits the first $k$ input signals unchanged, and inverts the last input signal *iff* the first $k$ inputs are all 1; clearly this input-output mapping is bijective. Figure 1 shows examples of *k*-CNOTs and the standard graphic symbols used for them. If $k = 0$, a *k*-CNOT is a simple NOT gate or inverter. A 1-CNOT gate is simply known as CNOT gate. A 2-CNOT gate is known as Toffoli gate. (NOT, CNOT, TOFFOLI) constitute NCT library. Since a Toffoli gate can implement the NAND function, any Boolean function can be implemented by a *k*-CNOT circuit. A k-CNOT gate can be implemented using 1 k-input AND gate and an EXOR gate as shown in Fig 2. Frequency of switching in a k-CNOT gate is very less, i.e. $1/2^k$. . Figure 3 shows k-CNOT implementation of a single ILA cell which can be connected as 1-D homogeneous array of identical cells to construct a reversible Ripple Carry Adder (RCA).

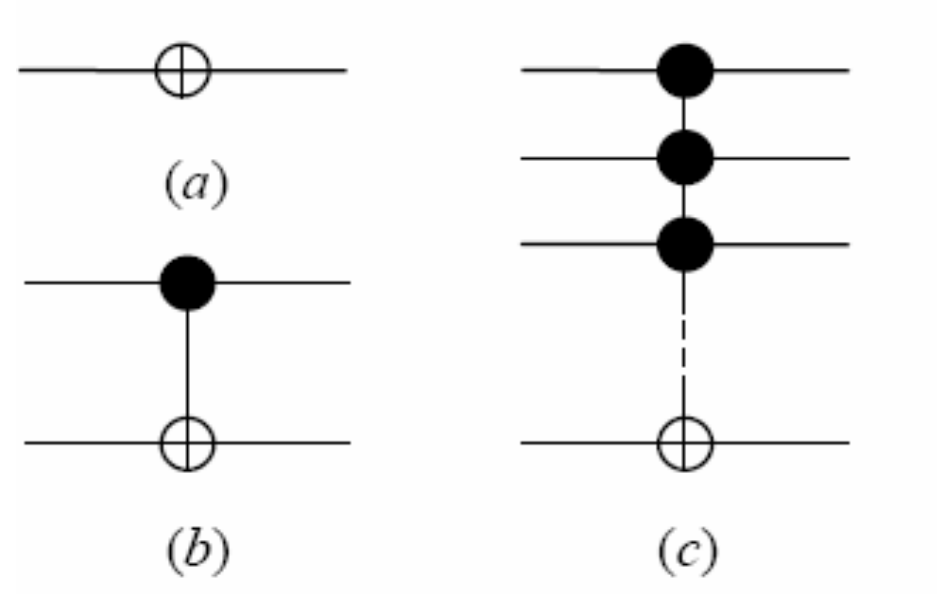

**Figure 1**: (a) NOT (b) CNOT and (c) general k-CNOT gate

The paper is organized as follows. In section 2 the assumed Single Cell Fault (SCF) model has been discussed. In section 3, how 1-D array of identical cells can realize a reversible Ripple Carry Adder (RCA) has been discussed. In section 4 C-Testability of ILAs has been proven and it has been shown that number of tests required is equal to number of entries in the cell truth table. In section 5 we generate tests for 1D ILAs. In section 6 we present test generation technique for 2D ILAs which can be actually extended for any n-dimensional ILAs. Then, Uniform-Testability, i.e. U-Testability has been defined and heterogeneous reversible ILAs have been shown to be U-Testable. In Section 7, we then examine the testability of ILAs under other fault models esp. Multiple Cell Fault (MCF) model. Section 8 concludes the paper.

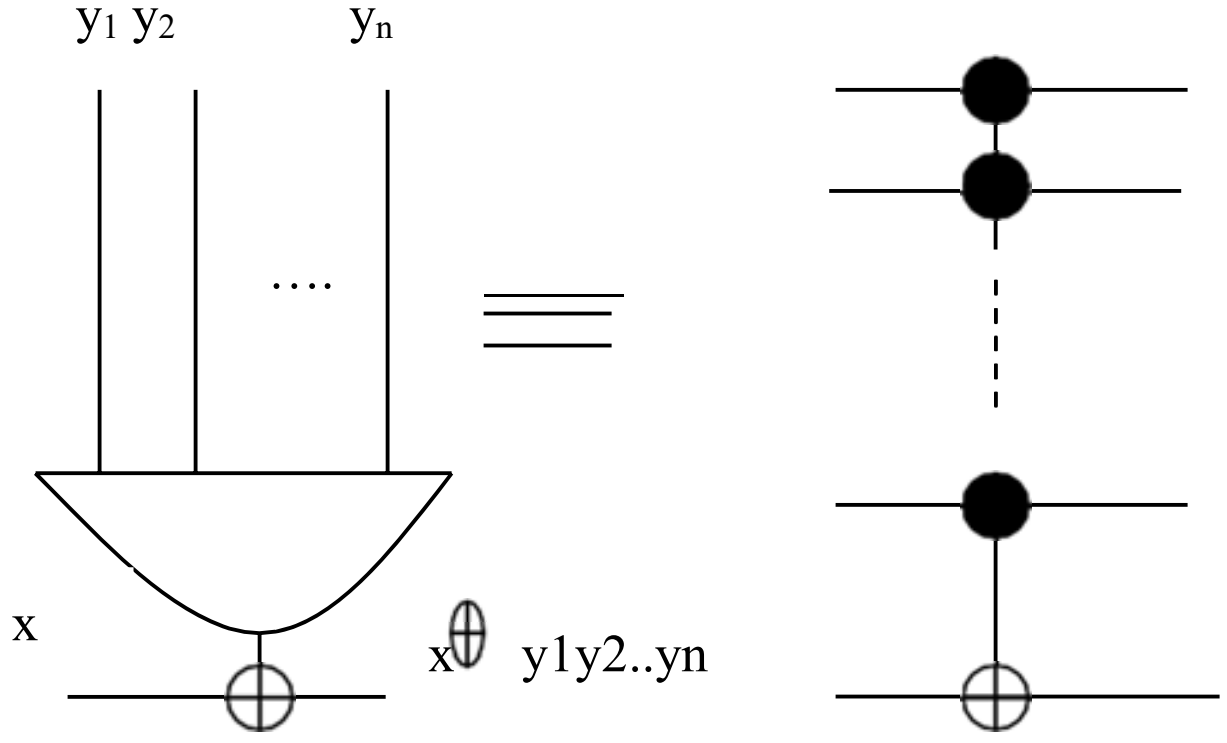

**Figure 2**: Implementation of k-CNOT gate

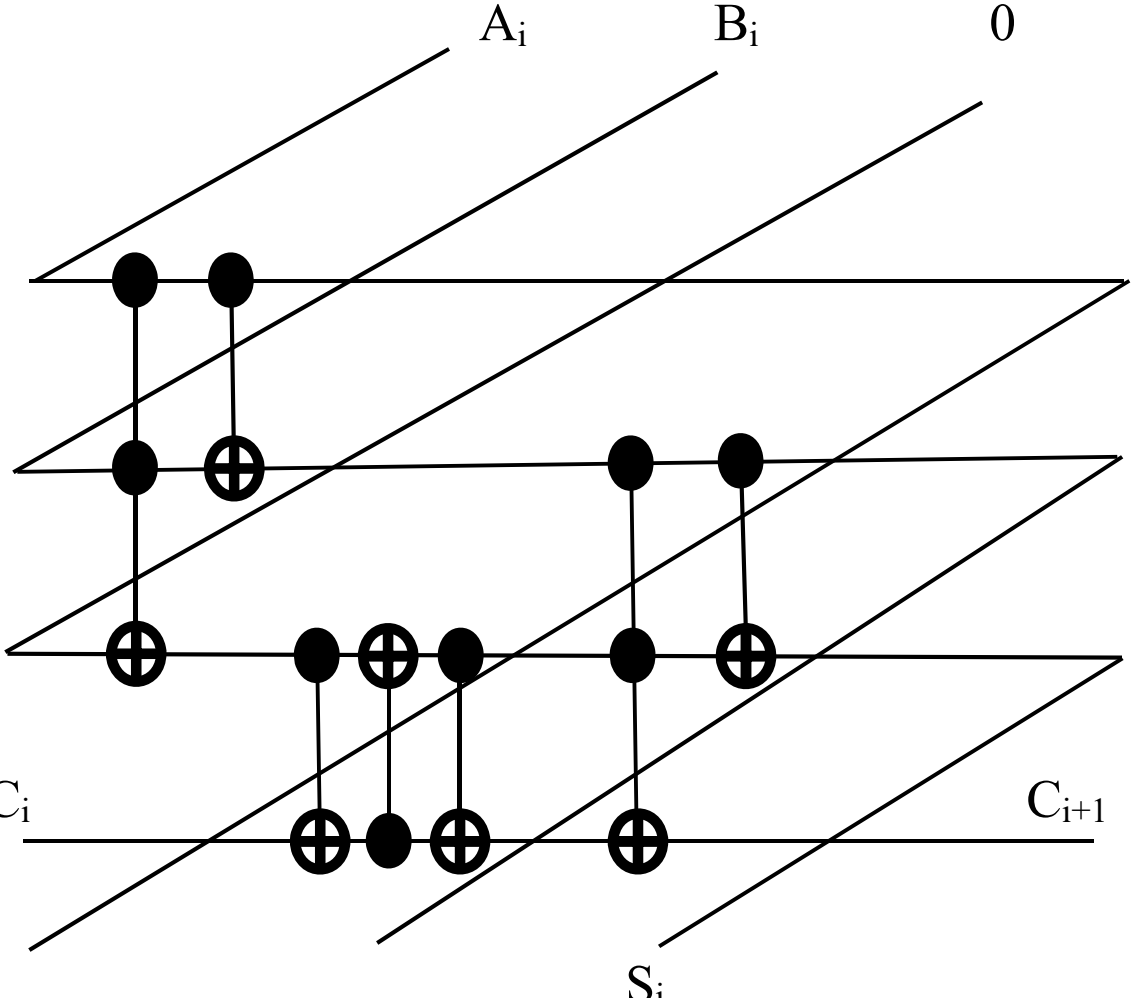

**Figure 3**: A (1,3)-ILA cell with 1 *h*-wire and 3 *v*-wires

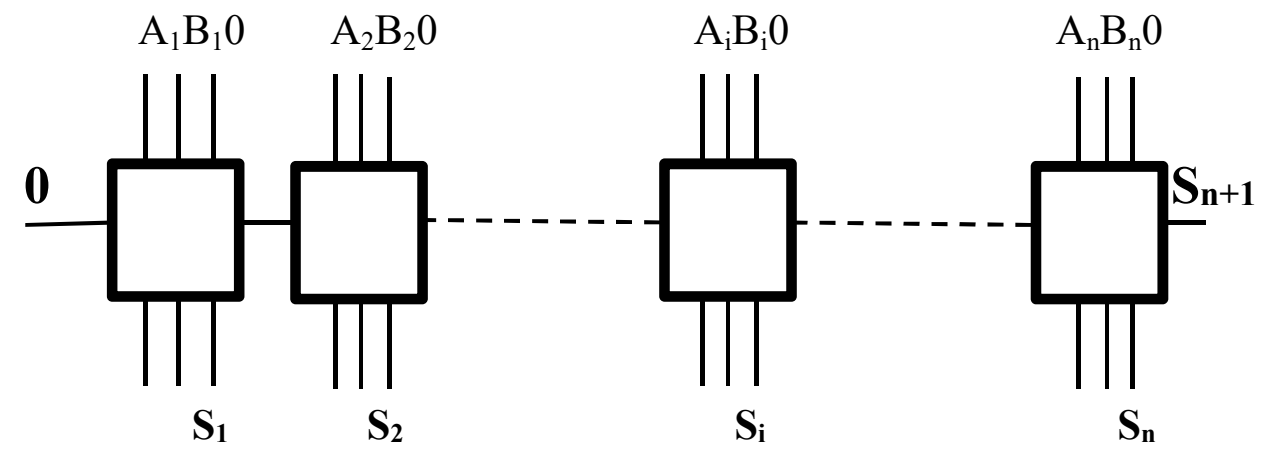

**Figure 4**: 1-D Array of (1,3)-cells realizing n-bit RCA

## II. THE FAULT MODEL

We assume that the truth table of a single ILA cell can be altered in any manner in presence of a fault as long as it remains combinational. In such a scenario all possible inputs must be applied to each of the ILA cells and we must ensure any such single cell fault (SCF) must propagate to the observable outputs of the ILA (for ex. in case of 2D ILA, vertical outputs of the last row and horizontal outputs of the last column).

## III. RIPPLE CARRY ADDER

ILA consists of identical cells arranged in a geometrically regular interconnection pattern. A cell which receives $h$ horizontal inputs $h_{in}$s, $v$ vertical inputs $v_{in}$s and produces $h$ horizontal outputs $h_{out}$s, $v$ vertical outputs $v_{out}$s is referred to as $(h, v)$-cell. See Fig 3. The $k_{max}$ of the $k$-CNOT gates that implement the internal circuitry of the ILA cell is bounded by: $k_{max} \leq (h + v - 1)$. Due to bijectivity it maps every distinct $(h_{in}, v_{in})$ to $(h_{out}, v_{out})$. We denote the cell function realized by $(h, v)$-cell as $f$ where:

$$f\colon \{0,1\}^h \times \{0,1\}^v \rightarrow \{0,1\}^h \times \{0,1\}^v$$

$f$ is clearly bijective. In ILA of one dimension the cells connected in a line. In 2D ILA, cells are interconnected in a rectangular structure. If $n$ (1, 3)-cells as shown in Fig 3 are connected as an 1D array (as shown in Fig. 4), it realizes an $n$-bit Ripple Carry Adder as each $S_i$ and $C_{i+1}$ realize sum and carry of 1-bit Full Adder respectively where:

$$S_i = A_i \oplus B_i \oplus C_i$$
$$C_{i+1} = A_iB_i \oplus B_iC_i \oplus C_iA_i$$

## IV. C-TESTABILITY OF ILAS

**Theorem 1**: All SCFs in one dimensional Reversible ILA of $p$ $(h, v)$-cells can be detected by $2^{(h+v)}$ tests independent of $p$.

**Proof**: As the cell function $f$ realized by $(h, v)$-cell is bijective (*one to one* and *onto*, i.e. both *surjective* and *injective*), any single cell fault propagates to either $h_{out}$ or $v_{out}$ of the cell in which the failure occurs. As all other cells are bijective and fault-free (SCF), this ensures that any SCF propagates to an observable output ($h_{out}$s of the rightmost cell or $v_{out}$s of all $p$ cells) for any vertical input $v_{in}$ to that faulty cell. This implies $2^{(h+v)}$ tests suffice to detect all SCFs.

**Theorem 2:** All SCFs in two dimensional Reversible ILA of $p$ rows and $q$ columns constructed from $(h, v)$-cells can be detected by $2^{(h+v)}$ tests independent of $p$ and $q$.

**Proof:** As the cell function f realized by $(h, v)$-cell is bijective; the fault propagates to either $h_{out}$s or $v_{out}$s of the faulty cell. If it propagates to $h_{out}$s, then the cell just at the right of the faulty one propagates the fault to its $h_{out}$s and $v_{out}$s (as it is assumed to be fault-free; only one cell can be faulty at a time); else the cell just beneath the faulty one propagates it to its $h_{out}$s or $v_{out}$s. This implies $2^{(h+v)}$ tests suffice to detect all SCFs in 2D ILAs.

Consider 1D and 2D ILAs constructed from the cell shown in Fig. 3. Both the ILAs can be tested for SCFs using only $2^{(1+3)}$ = 16 tests. Any ILA constructed from (1, 3)-cells can not be tested for SCFs by less than 16 tests. This testability is known as Optimal-Testability (i.e., O-Testability), where number of tests is independent of size of the ILA and equal to number of entries in the cell truth table. It can be proven that $2^{(h+v)}$ – testability implies following properties:

1. The $x$-transition diagram (transitions on $h$ wires) consists of edge disjoint Euler tours or Euler tours. See Fig. 5.

2. The $y$-transition diagram (transitions on $v$ wires) consists of edge disjoint Euler tours. See Fig. 6.

3. The state transition diagram (transitions on $h \mid v$, where | is concatenation operator) comprises of vertex-disjoint Euler tours or Hamiltonian tours. See Fig. 7. The corresponding $x$ and $y$ transitions also form edge-disjoint Euler tours in $x$ and $y$-transition diagram respectively.

4. The state transition (concatenation of transitions of all dimensions) of an $n$-dimensional ILA can be decomposed into vertex disjoint Euler tours or Hamiltonian tours and only one way such decomposition is possible.

5. The transition diagram for each dimension corresponding to decomposition of state table also consists of edge-disjoint Euler tours.

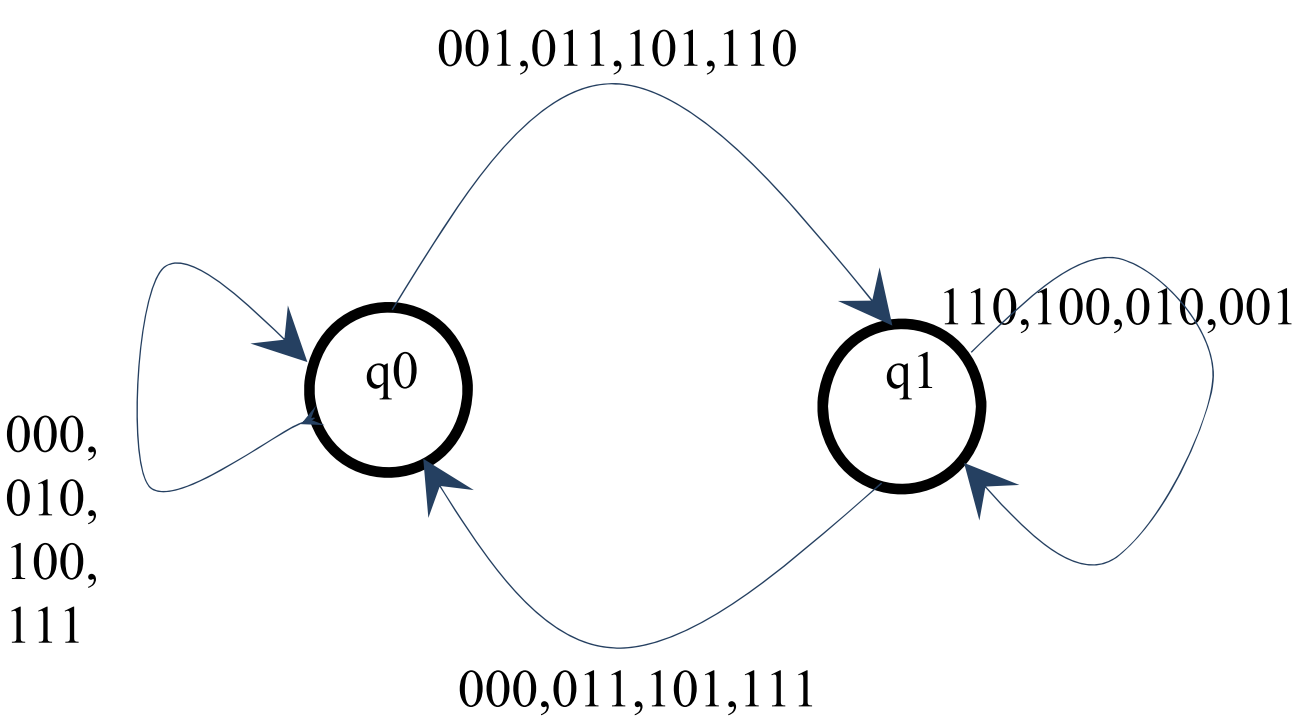

**Figure 5**: $x$-transition diagram for (1,3)-cell of Fig. 3

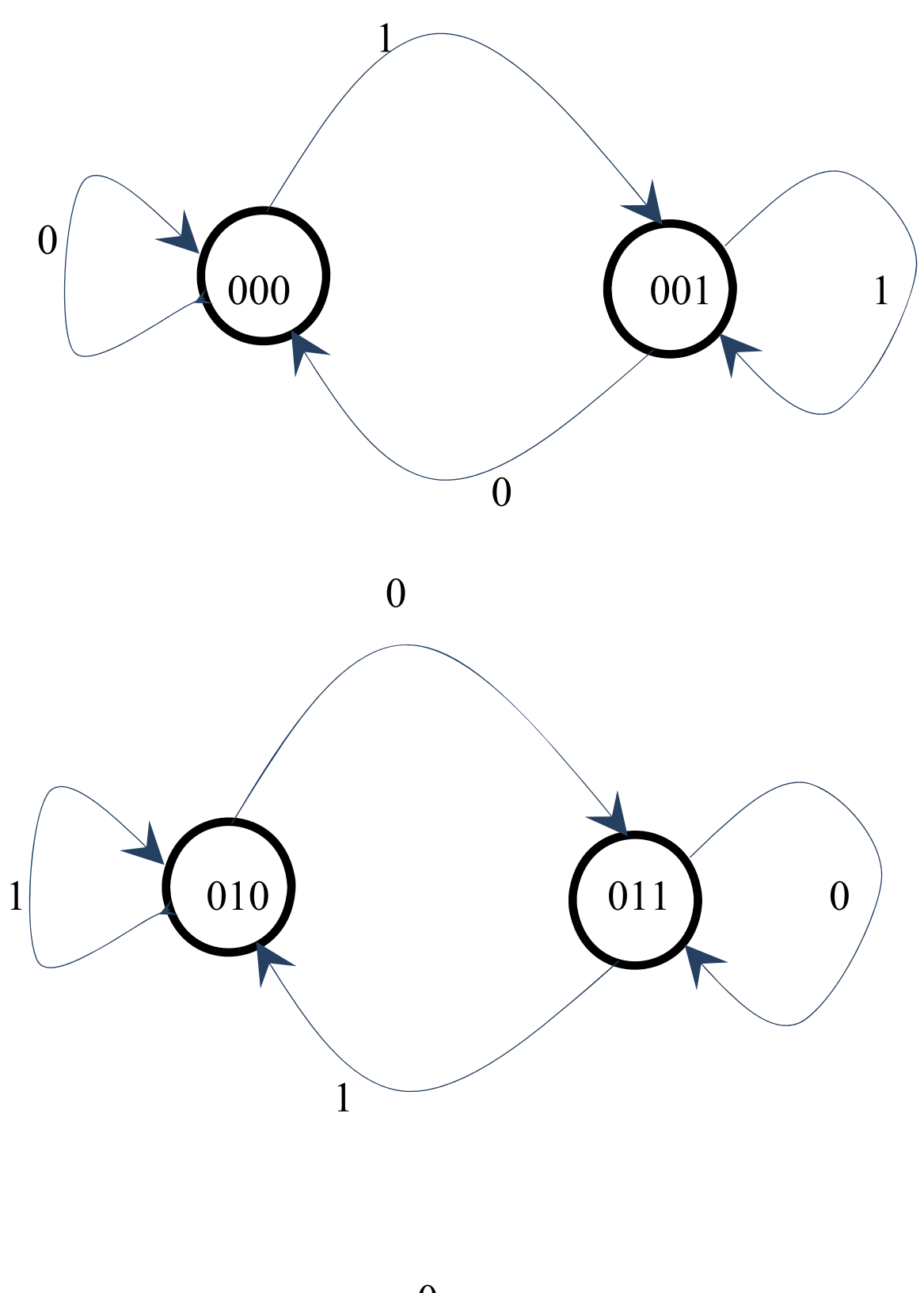

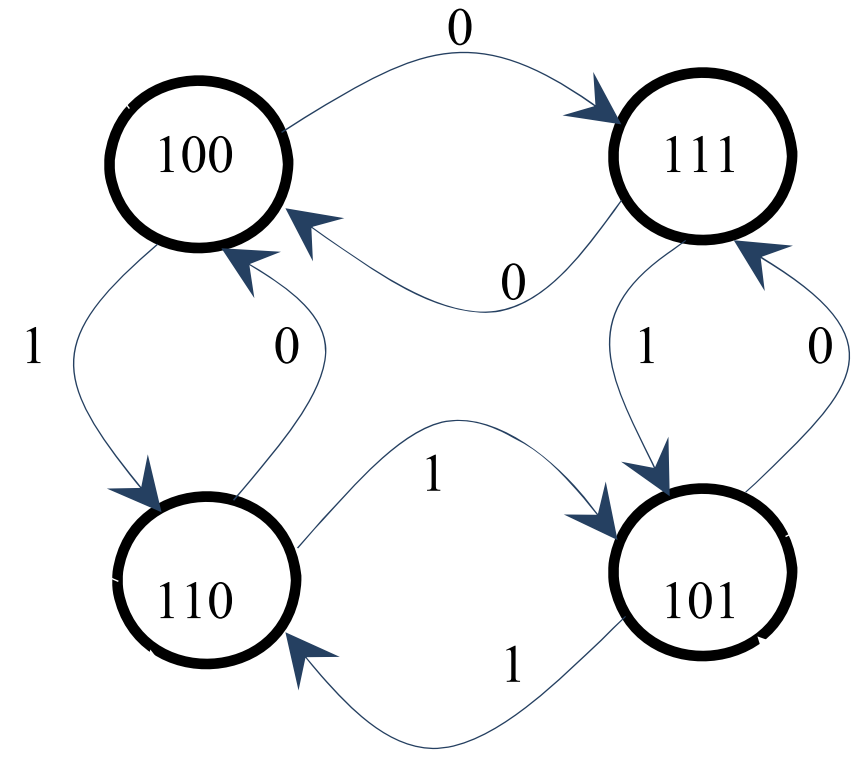

**Figure 6**: $y$-transition diagram for (1, 3)-cell of Fig. 3

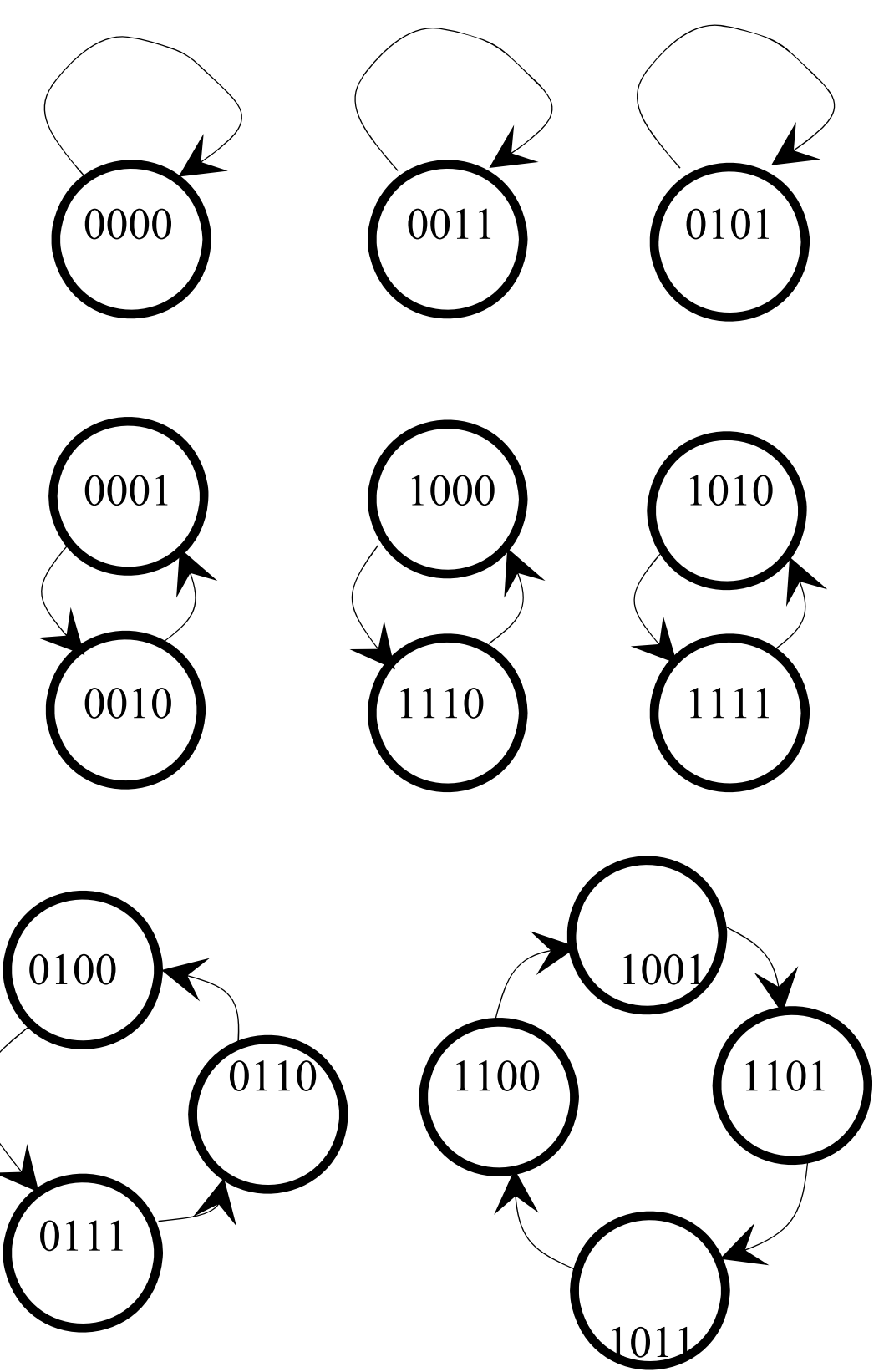

**Figure 7:** state transition diagram (v | h) for (1,3)-cell of Fig 3.

## V. One-Dimensional ILAS

We use property 1 to generate tests for 1D ILAs. Given, cell truth table we construct *x*-transition diagram and decompose it into edge-disjoint Euler tours. Please note that *x*-transition diagram can be decomposed in more than one ways and each of them gives a solution. We illustrate the test generation of 1D ILA constructed from (1, 3)-cells shown in Fig. 3. If we decompose the *x*-transition diagram of Fig. 5 as follows: {(q0, 000), (q0, 010), (q0, 100), (q0, 111), (q1, 110), (q1, 100), (q1, 010), (q1, 001), (q0,001)→(q1,000), (q0, 011)→ (q1,011), (q0, 101)→ (q1,101), (q0, 110)→ (q1,111)}, then we generate following 16 tests: $TS^{1D}_1$ applies (q0, 000) in all the cells; similarly $TS^{1D}_2$, $TS^{1D}_3$, $TS^{1D}_4$, $TS^{1D}_5$, $TS^{1D}_6$, $TS^{1D}_7$, $TS^{1D}_8$ apply (q0, 010), (q0, 100), (q0, 111), (q1, 110), (q1, 100), (q1, 010), (q1, 001) respectively in all the cells. $TS^{1D}_9$ applies (q0, 001) in odd cells and (q1, 000) in even cells. $TS^{1D}_{10}$ applies (q0, 001) in even cells and (q1, 000) in odd cells. Similarly $TS^{1D}_{11}$, $TS^{1D}_{12}$, $TS^{1D}_{13}$, $TS^{1D}_{14}$, $TS^{1D}_{15}$, $TS^{1D}_{16}$ alternate (q0, 011)→ (q1,011), (q0, 101)→ (q1,101), (q0, 110)→ (q1,111) along the row. The complexity of the algorithm is $O(E+V)$ ($E$= no. of edges = $2^{(h+v)}$ and $V$ = no. of vertices = $2^{(h+v)}$) = $O(1)$.

## VI. Two-Dimensional ILAs

2D ILAs constructed from ($h$, $v$) cells can be tested for SCFs by using $2^{(h+v)}$ tests. For the ILA cell of Fig. 3, 2D ILA built from this cell can be tested by 16 tests only. We can use property 1 in conjunction with prop. 2 to generate tests for 2D ILA. This implies that as the *x*-transition diagram can be decomposed into more than one ways, we need to find out one such decomposition for which the corresponding transitions also form closed walks in the *y*-transition diagram. Please note that for all possible decompositions of *x*-transition diagram, the corresponding transitions don't necessarily form closed walks in the *y*-transition diagram. But the problem of finding all possible decompositions of x-transition diagram into edge-disjoint Euler tours is *NP-hard*. Actually this can be overcome if we use the prop. 3 of state transition diagram (i.e. transitions on both *h* & *v* wires).This method of deriving tests for 2D ILA can be easily converted to an *ATPG* program which is based on the following corollary.

**Corollary 1:** As the state transition diagram can be decomposed into vertex-disjoint Euler tours or Hamiltonian tours and only one such decomposition exists, the corresponding transitions in *x* and *y* transition diagrams also form closed walks.

Consider the 8 cycles in the state transition diagram in Fig. 7. These cycles correspond to the following transitions:

1. $TS^{2D}_1$ applies (q0, 000) in all the cells of 2D ILAs
2. Similarly $TS^{2D}_2$ and $TS^{2D}_3$ apply (q1, 001) and (q1, 010) in all the cells.
3. $TS^{2D}_4$ and $TS^{2D}_5$ alternate (q1, 000) → (q0, 001) along the rows and columns both vertically and horizontally (odd and even).
4. Similarly $TS^{2D}_6$, $TS^{2D}_7$, $TS^{2D}_8$ and $TS^{2D}_9$ alternate (q0,100)→(q0,111) along the columns and (q0, 101)→(q1,111) along the rows and columns.
5. $TS^{2D}_{10}$, $TS^{2D}_{11}$ and $TS^{2D}_{12}$ alternate (q0, 010)→(q1, 011)→ (q0, 011) every 3 rows and columns of the 2D ILA.
6. Similarly $TS^{2D}_{13}$, $TS^{2D}_{14}$, $TS^{2D}_{15}$ and $TS^{2D}_{16}$ alternate (q0, 110)→ (q1, 100)→(q1,110) → (q1,101) every 4 rows and columns of the 2D ILA.

As cell truth table need not be analyzed for deriving test set for testing Reversible ILAs, it turns out that Reversible ILAs are more easily testable than what O-Testability characterizes. The test set generation algorithm has been outlined in Fig. 8. This kind of testability is being defined as *Uniform-Testability* or *U-Testability*. The test generation method need not consider the internal circuitry (state transition diagram) and any Reversible *n*-Dimensional ILA which can be tested with a Uniform Test Set. Interestingly, so far it has been discussed how Optimal-Test Sets can be derived for Reversible ILAs but the arrays have been assumed to be homogeneous, i.e. all the cells are identical. It can be shown that *n*-Dimensional Reversible Heterogeneous ILAs comprising of non-identical cells are U-Testable (also O-Testable and C-Testable). Hence, the theorem follows:

**Theorem 3:** *n*-Dimensional Reversible Heterogeneous Iterative Logic Arrays are Uniform-Testable (U-Testable) under SCF.

***Algorithm:***

*Consider a Heterogeneous Reversible n-Dimensional ILA comprising of n dimensions with*
*$(d_1, d_2, d_3,\ldots\ldots,d_n)$-cells.*

*For $i_1=0$; $i_1 < 2^{d1}$; $i_1$++)*
*{*
*Let $r_1$ be the binary representation of $i_1$*
*Apply $r_1$ along the dimension 1 at the boundary cells*
*For ($i_2=0$; $i_2 < 2^{d2}$; $i_2$ ++){*
*Let $r_2$ be the binary representation of $i_2$*
*Apply $r_2$ along the dimension 2 at the boundary cells*
*…………………………*
*…………………………*
*For($i_n=0$; $i_n < 2^{dn}$; $i_n$ ++){*
*Let $r_n$ be the binary representation of $i_n$.*
*Apply $r_n$ along the dimension n at the boundary cells*
*}*
*…………………………*
*…………………………*
*}*
*}*

**Figure 8:** Test Generation Algorithm for U-Testable ILAs

## VII. OTHER FAULT MODELS

As we are considering fault testing for ILAs, the more appropriate fault model would be the one where faults have been assumed to be manifested at the cell outputs. Though in recent past several fault models have been proposed and test generation under those fault models have been reported [10, 11, 12] but none of these takes into account the cellular structure of ILA like circuits while deriving the test set. If any other such fault occurs and the cell output is affected in some way, the SCF subsumes all such faults. The another fault model which considers testing for ILAs is Cell Delay Fault Model which addresses fault when the ILA implements sequential logic. It has been shown that 2D homogeneous array of bijective (*h, v*)-cells can be tested using $(h + v).2^{(h+v)}$ tests for cell delay faults [13]. Huang etal has investigated testable implementations of Quantum-dot-Cellular Automata (QCA) based one-dimensional reversible arrays under Multiple Cell Fault (MCF) model [14]. It has been shown that fault masking is unavoidable in presence of Multiple Faults in 1D array. They have also reported a technique for achieving C-Testability of 1D array by adding lines for controllability and observability. Consider a 2D $p\times q$ array of (*h, v*)-cells. Now let's assume that we can separately test each of the *p* rows as 1D array of *q* cells using *R(q)* tests according to techniques shown in [14] which is proportional to the number of cells in each row. Such characteristic can be achieved for a homogeneous 2D ILA of (*h, v*)-cells if following conditions of Column-Separability are satisfied [15]:

1. $h(a, y)=a$ *for all* $y \in \{0, 1\}^v$

2. $v(a, b) =b$ for some *b*, where *a* and *b* are fixed sequence.

This can be achieved by modifying the (*h, v*)-cell circuit by adding an extra wire as shown in Fig. 9, where *a=00* and *b=000*. The *x*-transition diagram of the modified (*h, v*)-cell consists of strongly connected components.

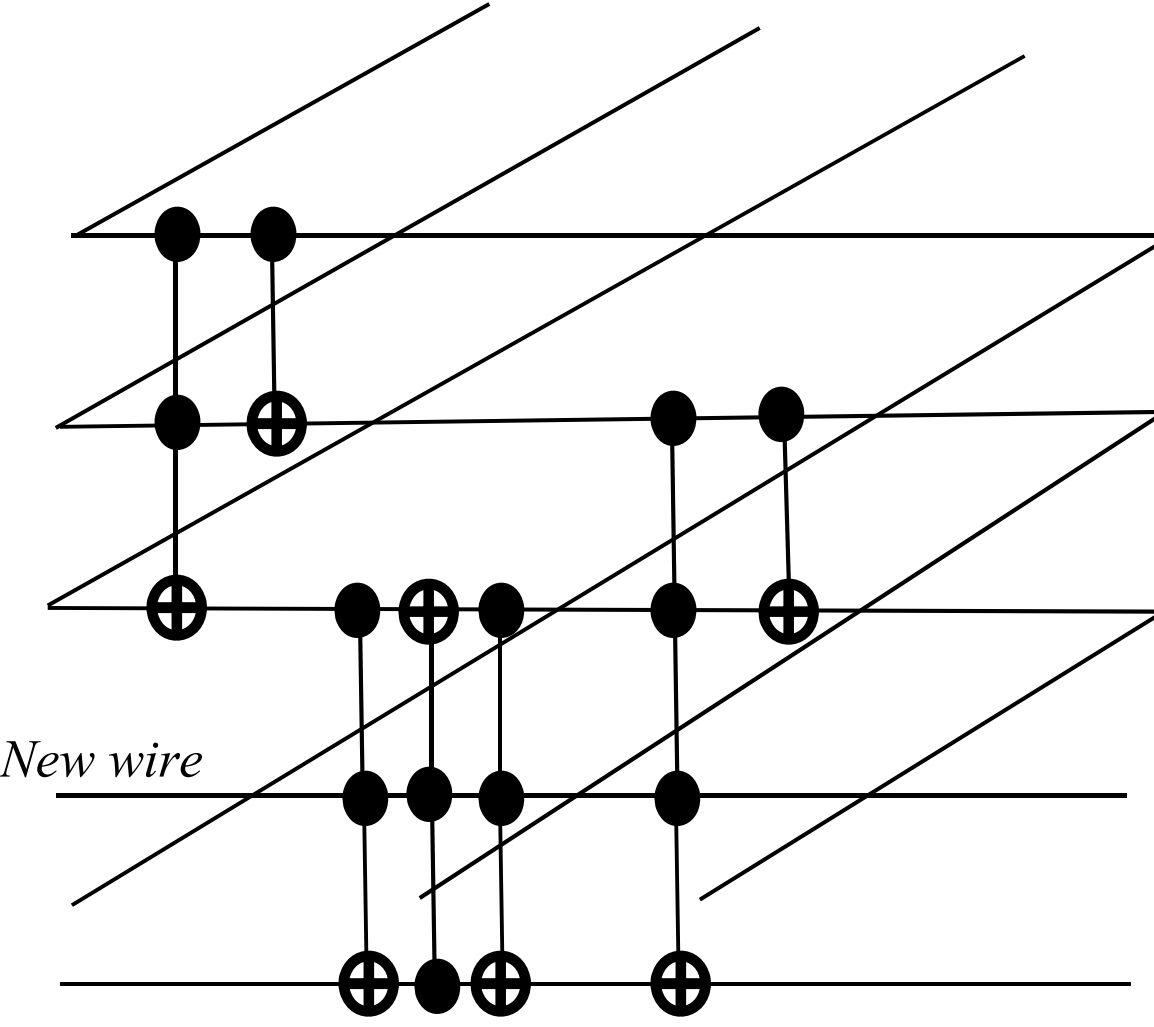

**Figure 9**: Column-separable Design of RCA cell

It has been shown that an entire $p\times q$ 2D array of column-separable *(h, v)*-cells can be tested for Multiple Cell Faults using $q(2^h -1) + p(2^v - 1) + p(R(q))$ tests.

## VIII. CONCLUSION

In this paper, we have characterized the testability properties of *n*D Reversible homogeneous and heterogeneous ILAs. As each cell is bijective and operations are linear (in *GF*(*2*)) it makes the array of reversible cells easily testable. With the emergence of 3D ICs, the idea of 3D ILA testability may be applied for testing 3D chips.